\documentclass{cimento}
\usepackage{graphicx}        
\usepackage{color}           
\usepackage{url}             

\title{The power spectrum of the seeing during solar observations}

\author{C.~Sigismondi\from{ins:s}\from{ins:i}\from{ins:a}\from{ins:u},
        A.~Raponi\from{ins:s}\ETC,
        X.~Wang\from{ins:e},
        G.~De Rosi\from{ins:s},
        M.~Bianda\from{ins:a},
\atque
        R.~Ramelli\from{ins:a},
        }

   \instlist{\inst{ins:s} Sapienza University of Rome, P.le Aldo Moro 5 00185, Roma (Italy) 
              
             \inst{ins:i} ICRA, International Center for Relativistic Astrophysics, Roma 
                    
             \inst{ins:a} Istituto Ricerche Solari di Locarno, Switzerland
                     
             \inst{ins:u} Universit\'e de Nice-Sophia Antipolis/Fizeau, France
                    
             \inst{ins:e} Huairou Solar Observing Station, National Astronomical Observatories, Chinese Academy of Sciences, Beijing, China                
             }
\PACSes{\PACSit{95.10.Jk}{Astrometry and reference systems}
\PACSit{92.60.hk}{Atmospheric turbulence}
\PACSit{96.60.-j}{Solar physics}
\PACSit{96.60.Bn}{Solar diameter}}

\begin{document}

\maketitle

\begin{abstract}
Measurements of the power spectrum of the atmospheric seeing in the line of sight of the Sun, in the range 0.001-1 Hz, have been performed in Santa Maria degli Angeli Lucernaria Dome, at IRSOL and in Huairou Station. This study is aimed to understand the criticity of the meridian transits method for solar diameter monitoring.
\end{abstract}

\section{Measuring the solar diameter with drift-scan method}

The method consists in the use of a fixed telescope and the observation of the drift of the solar image through a meridian or a given almucantarat. Knowing with accuracy the speed of transit of the Sun and measuring the transit time, it can give an accurate measure of the diameter. Should also be noted that the measured time is not affected by atmospheric refraction. The advantage of drift scan method is that it is not affected by optical defects and aberrations, because both edges are observed aiming in the same direction, moreover with parallel transits one can gather N observations, at rather homogeneous seeing conditions, during the time of a measurement.
The measurements of the solar diameter by meridian transit were monitored on a daily basis since 1851 at Greenwich Observatory \cite{Gething} and at the Campidoglio (Capitol) Observatory in Rome since 1877 to 1937 (see Fig. 1).
Despite the advantages of this method, independent measurements seem to be inconsistent \cite{Wittmann}. This discrepancy could be due to the importance of seeing for periods longer than the typical time of the transit (about 2 minutes). 
To evaluate this hypothesis, we perform a survey on the power spectrum of the seeing.

\begin{figure}
\centerline{\includegraphics[width=1.0\textwidth,clip=]{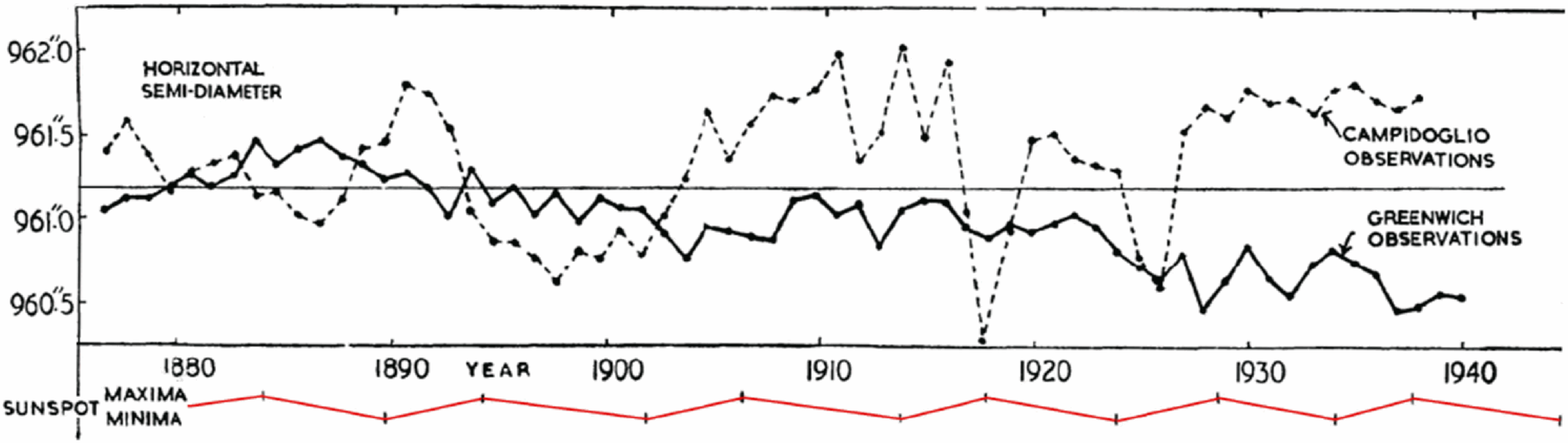}}
\caption{Horizontal semi-diameters of the Sun measured at Greenwich with Airy's meridian circle. Campidoglio observations are superimposed (Gething 1955). Straight lines correspond to a radius of 961.2 arcsec. 
The inconsistency between two data set is evident.
}
\label{Fig. 1}
\end{figure}

\section{Lucernaria  Discrete Power Spectrum}

The most simple way to measure the seeing is by projection of the solar image on a regular grid during a drift-scan observation. A videocamera records the transit of the solar limbs above the grid, and the time intervals required to cover the evenly spaced intervals of the grid are measured by a frame by frame inspection.
The standard deviation of these time intervals $\sigma$ [s] is related to the seeing $\rho$ [arcsec] by the approximate formula: $\rho = \sigma \cdot 15\cdot \cos(\delta_{\odot})$ where $\delta_{\odot}$ is the declination of the Sun at the moment of the observation.

The diffraction is the lower limit of the detectable amplitude of the seeing. 
For our application we used the Lucernaria Dome indoors Basilica Santa Maria degli Angeli e dei Martiri in Rome. The Lucernaria lenses are fixed solar telescopes with opening 6.3 cm and focal lenght 20 m \cite{Cuevas}. The resulting diffraction is 2.31 arcsec for $\lambda = 550$ nm.

Results in Fig. 2.

\begin{figure}
\centerline{\includegraphics[width=0.6\textwidth,clip=]{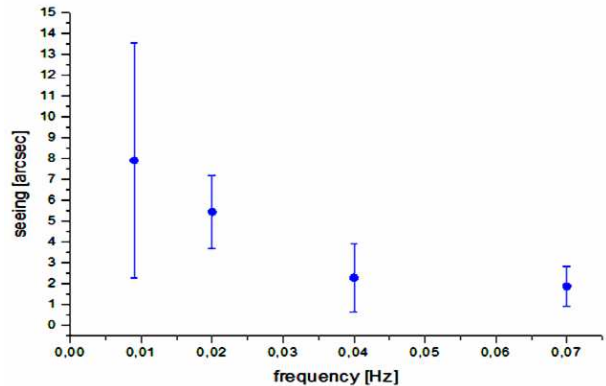}}
\caption{Power spectrum of the seeing: from the graphic is evident that there is significant power even at low frequencies.}
\label{Fig. 2}
\end{figure}

\section{IRSOL}
\subsection{Telescope}
The observation of the full solar disk is performed with Gregory-Coud\'e vacuum telescope 45 cm aperture and 25.0 m focal length.
The instrument gives a portion (about 100 arcsec x 200 arcsec) of the solar image, from wich we recover the curvature of the limb and the solar center.
\subsection{Detector}
The image of the Sun is projected on the CCD Baumer camera, and it is digitalized. The figures here represent the motion of the center of the Sun as recovered from the Northern limb, tracked for 1000 s, and the corresponding FFT power spectrum over frequencies, in abscissa, from 1 to 1/100 of such timespan.

Results in Fig. 3.

\begin{figure}
\centerline{\includegraphics[width=0.6\textwidth,clip=]{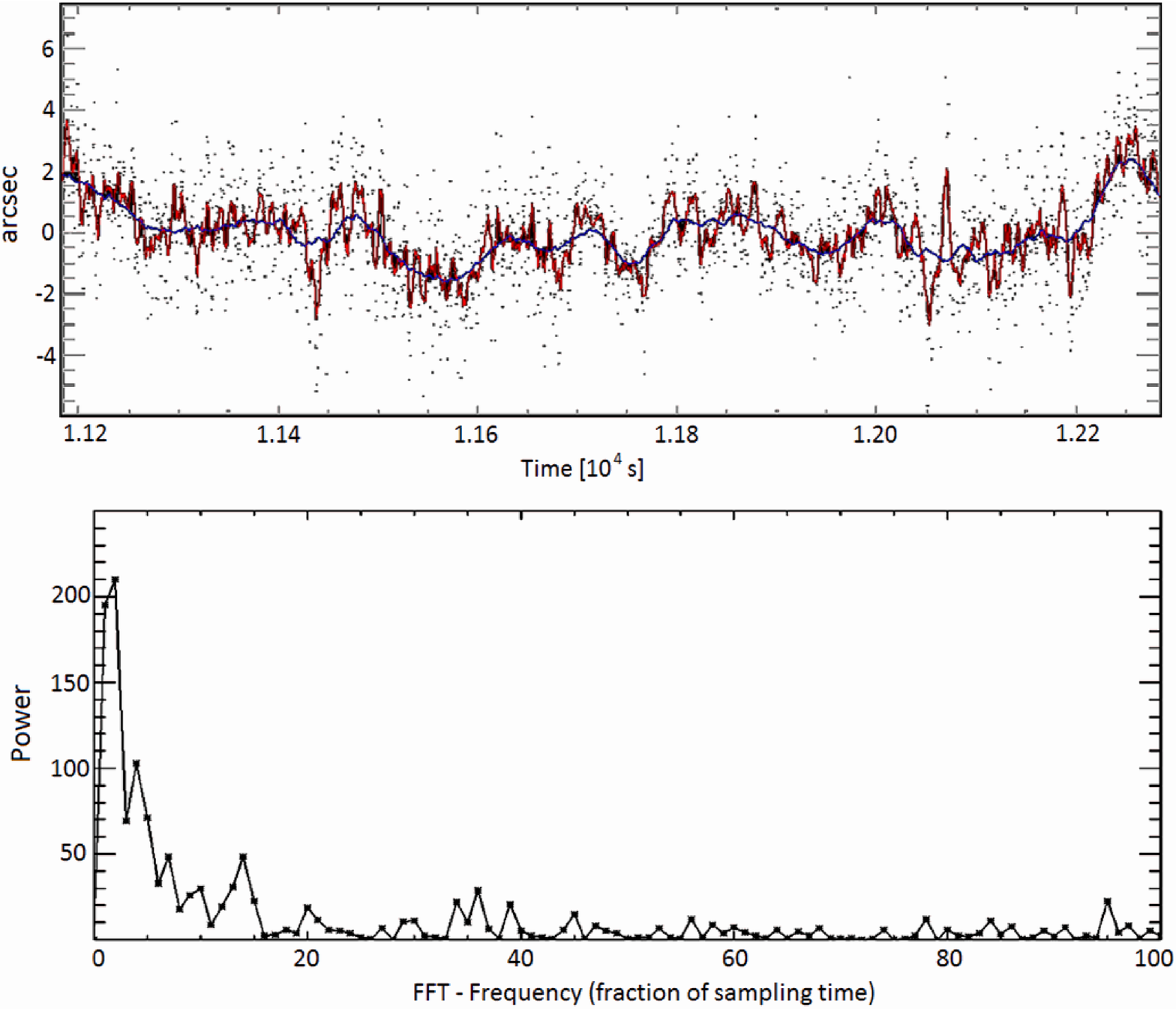}}
\caption{The Northern limb is considered. In this image, the effect of "sub-Hertz" fluctuations is evidenced to the scale of one arcsecond. This can explain the difference between two following measurements of the solar diameter with drift-scan method, and this also explain the need of several measurements to be statistically averaged in order to give a reference value for the solar diameter. But several measurements can occur under different meteorological conditions even in the same day, and the single measurement cannot be considered as statistically independent, as the gaussian hypothesis requires. That could be one of the reason of the great scatter that the yearly averages published for Greenwich and Capitol Observatory both show. }
\label{Fig. 3}
\end{figure}

\section{Huairou Solar Station}
\subsection{Telescope}
The observation of the full solar disk is performed with the Solar Magnetism and Activity Telescope (SMAT) \cite{Zhang} that is a telescope with a tele-centric optical system of 10 cm aperture and 77.086 cm effective focal length realized to investigate the global magnetic configuration and the relationship with solar activities. 
The birefringent filter for the measurement of vector magnetic field is centered at 532.419 nm (Fe) and its bandpass is 0.01 nm.
\subsection{Detector}
A CCD camera, Kodak KAI-1020, is used for the measurement of full disk. The image size of the telescope is 7.4 mm × 7.4 mm, and the size of CCD is 992 × 1004 pixels. 
The frame rate of the CCD camera is 30 frame $s^{-1}$ and its maximum transmission rate is 60 Mbyte $s^{-1}$. 
Two data collecting mode are performed with 2ms exposure time: 

 - 21 ms readout, 847 frames 
 
 - 7 s readout, 128 frames

Results in Fig. 4.

\begin{figure}
\centerline{\includegraphics[width=0.9\textwidth,clip=]{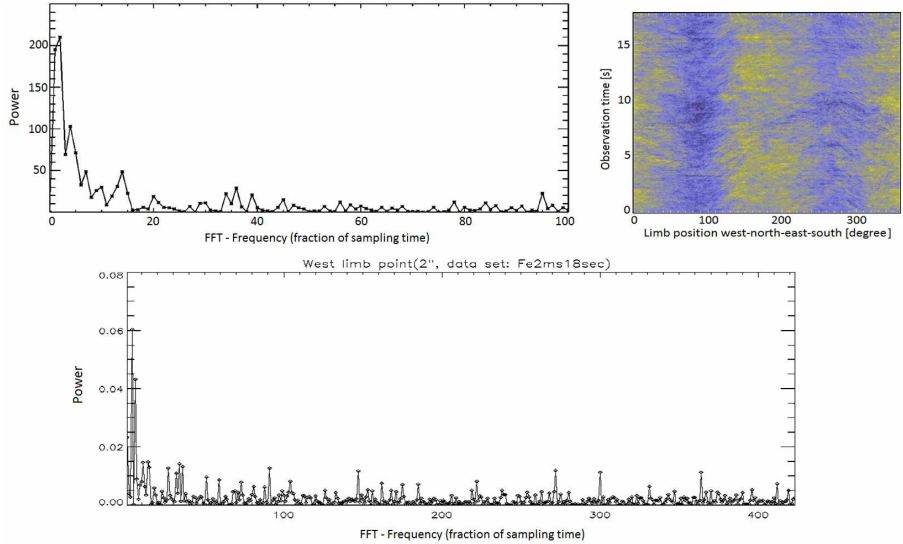}}
\caption{Fast Fourier Transforms are performed for both data collecting mode (upper left and lower panels). 
Differences between them lead us to consider that comparison must be made with care. 
In upper right panel the distance of the limb points with respect to the center of the Sun are in function of the angle and the time. The color is a measure of the distance.}
\label{Fig. 4}
\end{figure}

\subsection{Full disk radial pulsation analysis}
If we are interested in monitoring the radius of the Sun integrated over the solar disk, we must know how much the seeing affects this measure over time.

Results in Fig. 5.

\begin{figure}
\centerline{\includegraphics[width=0.7\textwidth,clip=]{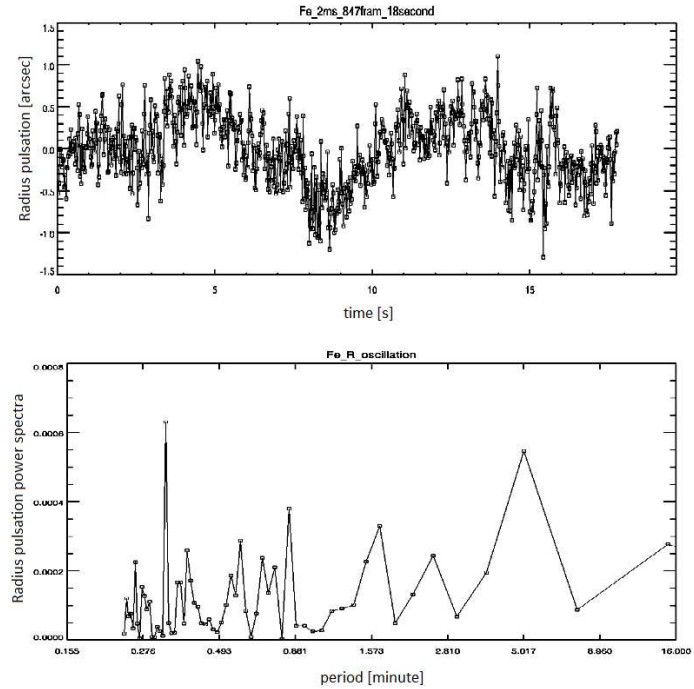}}
\caption{For every frame of the image we performed a fit with the inflection points position. 
The fit returns an average value of the radius as well as the position of the center of the Sun. We plot the evolution of r – r' as a function of time (upper panel), beeing r the raius obtained by each frame, and r' the average radius obtained by all frames. 
We obtain the power spectrum of the fluctuations through a Fast Fourier Transform (lower panel).}
\label{Fig. 5}
\end{figure}

\section{Conclusion}
The role of seeing fluctuations under 1/10 and 1/100 of Hertz is crucial in drift-scan measurements of the solar diameter, either meridian transits or almucantarat transits. This study firstly evidenced this effect in a clear
way. The fluctuation's scale is not defined here, since we did not apply the analysis to a single point only, but also on solar limb arches of several arcseconds (from 200 for IRSOL to the whole disk for Huairou Solar Station). Consequently the scale of the seeing (composed by blurring + image deformation + image motion) depends also on the algorithm used to define the solar figure.

An average made with N points distributed over all $360^{\circ}$ of position angles (Huairou Solar Station), is different by the one made over $12^{\circ}$ (IRSOL), and by the one, discrete, made on the preceding and following limb at Santa Maria degli Angeli Lucernaria by visual inspection (about $20^{\circ}$ of the solar limb involved). 
All these measurements have in common the detection of significan energy at the low frequencies regions around 0.01 Hz of the power spectrum. 
These effects have to be taken into consideration for a fruitful analysis of the solar diameter measured with these methods.

A full Sun imager would help to monitor the image motion occurred during a single transit observed with the drift-scan method. 


\bibliography{PSS}
   
\bibliographystyle{varenna}

\end{document}